\documentstyle[sprocl]{article}

\input{psfig}

\def\eg{{\it e.g.}}
\def\etal{{\it et al.}}
\def\etc{{\it etc.}}
\def\ie{{\it i.e.}}
\def\spose#1{\hbox to 0pt{#1\hss}}
\def\ltsim{\mathrel{\spose{\lower.5ex\hbox{$\mathchar"218$}}
     \raise.4ex\hbox{$\mathchar"13C$}}}

\begin{document}

\title{\uppercase{Ellipticals and bars: \\ central masses and friction}}

\author{J. A. SELLWOOD}

\address{Department of Physics \& Astronomy, Rutgers University \\ 136 
Frelinghuysen Road, Piscataway, NJ 08854-8019, USA \\E-mail: 
sellwood@physics.rutgers.edu} 

\maketitle\abstracts{I give a very brief review of aspects of internal dynamics 
that affect the global shape of a galaxy, focusing on triaxiality, bars and 
warps.  There is general agreement that large central masses can destroy 
triaxial shapes, but recent simulations of this process seem to suffer from 
numerical difficulties.  Central black holes alone are probably not massive 
enough to destroy global triaxiality, but when augmented by star and gas 
concentrations in barred galaxies, the global shape may be affected.  Even 
though we do not understand the origin of bars in galaxies, they are very useful 
as probes of the dark matter density of the inner halo.  Finally, I note that 
dynamical friction acts to reduce a misalignment between the spin axes of the 
disk and halo, producing a nice warp in the outer disk which has many of the 
properties of observed galactic warps.}

\section{Introduction}
I was asked by the organizers to discuss the huge topic of what $N$-body 
simulations can tell us about the shapes of galaxies.  As a comprehensive survey 
would be voluminous, I here restrict myself to a few internal evolutionary 
processes affecting galaxy shapes where there have been noteworthy recent 
developments.  I also point out some areas where a lot more work is required.  
Other speakers describe aspects of galaxy formation, and the properties of 
collapsed halos.

\section{Triaxiality and central masses}
Soon after Schwarzschild (1979) showed that a large fraction of the stars must 
pursue box orbits to support a triaxial mass distribution in an elliptical 
galaxy, it was realized that a central mass, or just a density cusp, might 
destroy triaxiality.  The reason, most clearly articulated by Gerhard \& Binney 
(1985), is that a box orbit carries a star arbitrarily close the center, where a 
central mass will deflect it through a large angle, scattering the star into an 
orbit of different shape.  The loss of stars on the structure-supporting box 
orbits must change the global shape of the galaxy, at least in the inner 
regions.  Only tube orbits, which avoid the center, remain once the potential 
becomes axisymmetric and the central mass concentration will no longer drive 
evolution.  This general expectation has been confirmed in $N$-body simulations, 
beginning with the pioneering work by May, van Albada \& Norman (1985) and most 
recently by Merritt \& Quinlan (1998, hereafter MQ98).

\begin{figure}[t]
\centerline{\psfig{figure=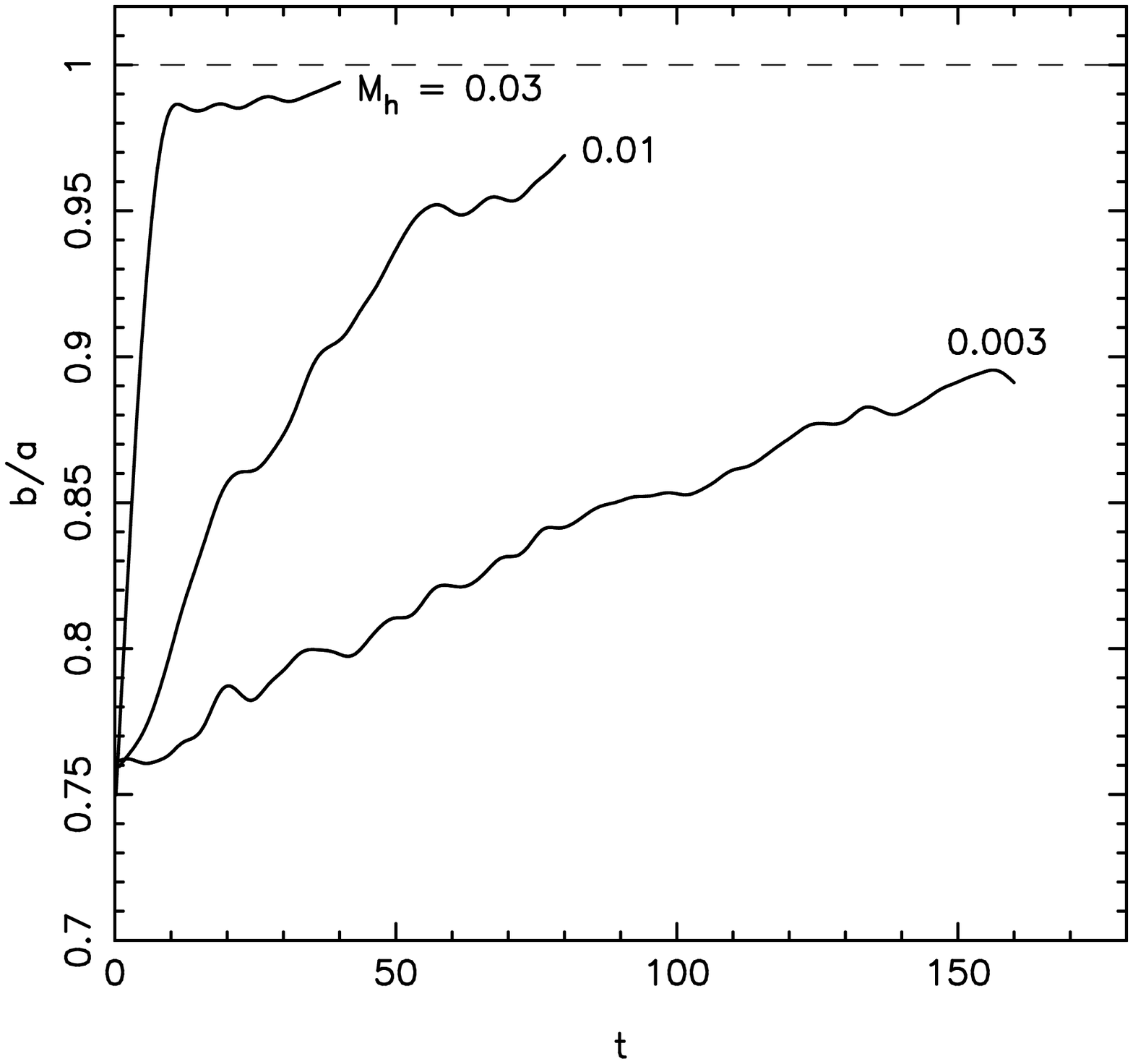,width=0.48\hsize,angle=0,clip=} \hfil
\psfig{figure=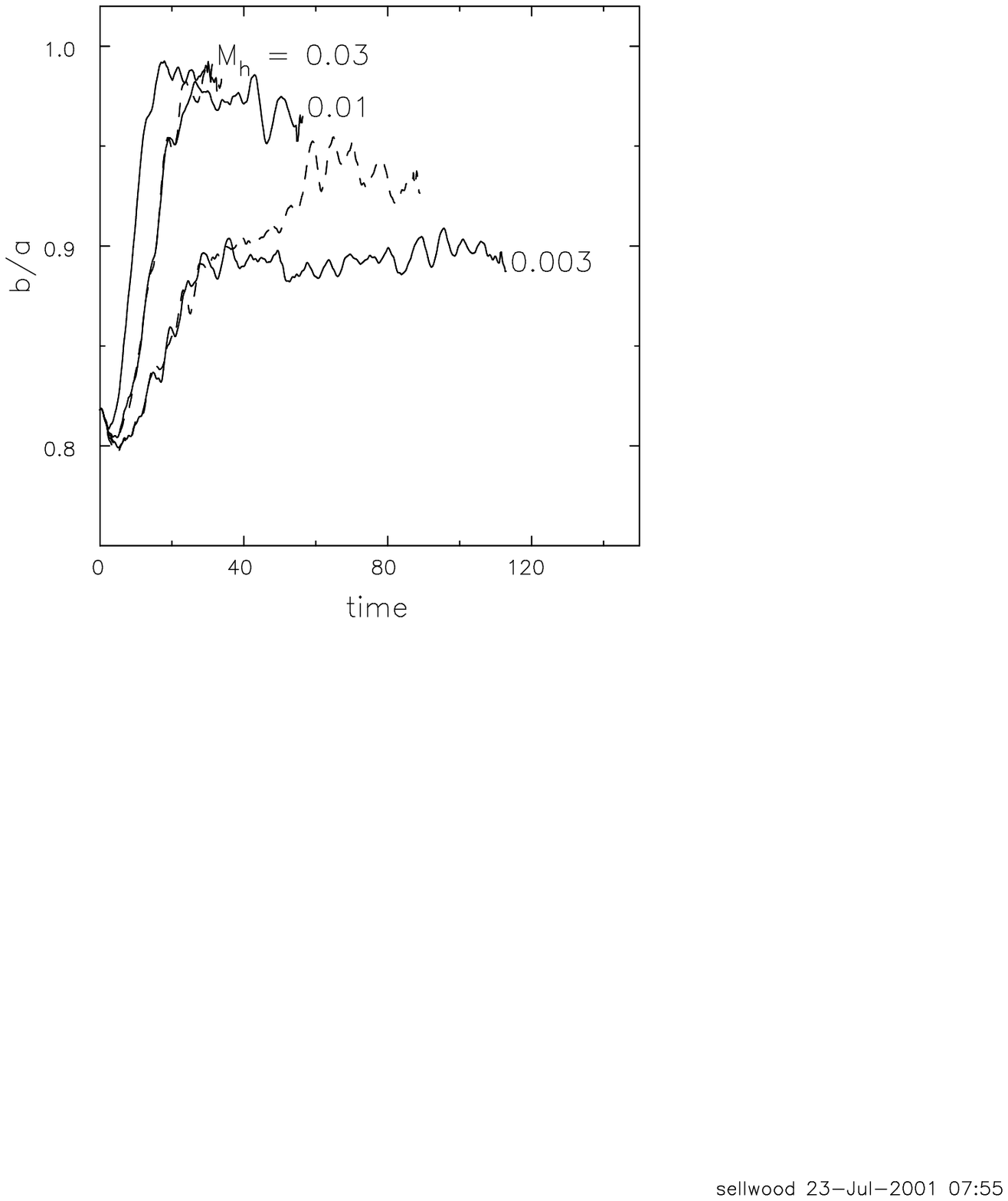,width=0.48\hsize,angle=0,clip=}}
\caption{Comparison between my results (right) and those obtained by MQ98 
(left).  The plotted quantity is the ratio of the intermediate- to long-axes of 
the moment of inertia tensor of the 50\% most bound particles as a function of 
time.  The three solid curves show the results from models into which three 
different central masses were grown over the time interval 0 to 15 for MQ98 and 
0 to 28 in my models.  The dashed curves on the right shows results from reruns 
of two cases using the SCF method.}
\end{figure}

While the simulations by MQ98 appear to have been performed with considerable 
care, one aspect of their results seemed highly surprising.  As shown in Figure 
1 (left), the shape of their initially triaxial model responded gradually to the 
introduction of a small central mass, and continued to evolve for many tens of 
dynamical times after the central mass reached a steady value at $t=5$ for M$_h 
= 0.3$ and $t=15$ for the two lower mass cases.  I suspected that this behavior 
was a numerical artifact, possibly caused by their use of the so-called SCF 
method proposed by Hernquist \& Ostriker (1992): although the basis set of 
functions is complete, truncation to a comparatively few terms provides a severe 
restriction on the complexity of shapes which can be represented.  Perhaps their 
numerical technique couples shape changes at different radii, which have greatly 
differing orbital time scales, rendering their simulations unable to sustain a 
shape which is axisymmetric near the center while remaining non-axisymmetric in 
the outer parts.  I therefore decided to check their work by following their 
procedure as closely as possible but employed a different method to compute the 
gravitational field.  The code, which closely resembles that previously 
described by McGlynn (1984) and requires somewhat less computer time than the 
SCF method (Sellwood 1997), expands the mass distribution in surface harmonics 
on a 1-D grid of shells, thus avoiding any coupling between the shape of the 
gravitational field at different radii.

My results (Figure 1, right) are markedly different from those reported by MQ98 
(Figure 1, left).  I find that the original triaxial shape both evolves rapidly 
as the central mass rises and stops evolving as soon as the increase stops at 
$t\sim 28$.  The rate of evolution in the two lower central mass models is 
clearly slower in the simulations by MQ98, whereas I find even more rapid 
evolution (not shown) when I grow the central mass as fast as did MQ98.

Reverting to an independent implementation of the SCF method with the same 
number of terms gives the results shown by the dashed lines in Figure 1 (right). 
 These lines track the solid curves as the central mass rises, but the later 
diverging trend, most obvious for the smallest central mass, shows that indeed 
the shape to continues to evolve after the central mass stops growing when the 
SCF code is used.  This behavior confirms my suspicions that the SCF method to 
compute the gravitational field caused the slow evolution reported by MQ98.  I 
have made a number of other tests, varying the usual numerical parameters of 
time step, softening length of the central mass, \etc, which have made no 
difference to the behavior.   While the slower initial evolution reported by 
MQ98 remains unexplained, the continuing shape changes in their models appear to 
result from a numerical artifact.

Black hole masses may be smaller than first believed (\eg\ Merritt \& Ferrarese 
2001) and their host galaxies therefore able to remain strongly triaxial.  I 
find that masses $\ltsim 0.1$\% have a negligible effect on the global shape for 
as long as I run the calculations.

The effect of a central mass on rapidly-rotating, triaxial bar is perhaps even 
less well understood.  The mass of a central black hole alone is unlikely to be 
large enough to affect the global shape, but augmented by a stellar cusp and 
other concentrations of stars and gas, the central masses can be enormous.  
There is general agreement that large central masses will destroy a bar; a mass 
of 5\% of that of the disk is clearly plenty (Norman \etal\ 1996), but Friedli 
(1994) claims that as little as 1\% is sufficient, while Sellwood \& Moore 
(1999) were able to grow a bar in a disk with a central mass larger than this!  
The criterion seems to be more complicated than just a simple mass fraction and 
the central density required to destroy a bar is not yet known.

\section{The Origin of Bars}
It is now clear that we do not understand the origin of bars in galaxies.  The 
original prejudice (\eg\ Sellwood \& Wilkinson 1993) that they were probably 
created by the well-known global instability (Miller \etal\ 1970; Hohl 1971; 
Kalnajs 1972) now seems untenable for two reasons.

First, most bright galaxies should be stable because they have dense bulges.  My 
own recent work (Sellwood 1985, 1999; Sellwood \& Evans 2001) has established 
beyond doubt the correctness of Toomre's (1981) idea that an entire disk can be 
stable if it has a dense center.  A disk with a rotation curve that stays high 
close into the center, even if fully-self-gravitating, has no tendency to form a 
bar; early simulations simply lacked the dynamic range in density and 
time-scales to show this.  However, a problem now arises for barred galaxies 
because they too have inner mass distributions which ensure a strong inner 
Lindblad resonance (ILR) that should have prevented the bar from forming.  
Indications of strong ILRs in nearby barred galaxies include nuclear rings (Buta 
\& Crocker 1993; Benedict \etal\ 1999), ring-like concentrations of molecular 
gas (\eg\ Sakamoto \etal\ 1999; Jogee \& Kenney 2000) and offset dust lanes 
(Athanassoula 1992).  Binney \etal\ (1991), Weiner \& Sellwood (1999) and others 
also find evidence for an ILR in the bar of the Milky Way, which seems 
inevitable given the central density distribution (Becklin \& Neugebauer 1968).  
This abundance of evidence is strongly suggestive, but indirect.  Definitive 
verification requires detailed modeling of the mass distribution and velocity 
field within the galaxy, together with a determination of the pattern speed in 
each case (Lindblad, Lindblad \& Athanassoula 1996; Regan, Vogel \& Teuben 1997; 
Weiner \etal\ 2001) -- all three cases reveal clear ILRs.  The question raised 
by these observations is that if the bars have strong ILRs, which we now know 
should have inhibited the bar instability, how did the bars form in the first 
place?  It is possible the central density rose after the bar formed, but the 
mass inflow required to produce such strong ILRs may well have destroyed the 
bar.

Second, bars seem to have formed long after their host disks came into 
existence.  There appears to be deficiency of bars in galaxies at $z>0.5$ (van 
den Bergh \etal\ 1996; Abraham \etal\ 1999), suggesting that bars form late in 
the evolution of a disk galaxy.  While some bars could be missed because of 
bandshifting (\eg\ Bunker \etal\ 2000), a marked deficiency remains (van den 
Bergh \etal\ 2000, 2001).  Thus most bars appear to form long after their host 
disks were assembled, which requires a slow, secular process quite different 
from the rapid dynamical instability.

I have observed an alternative mode of bar formation in number of simulations 
which remains consistent with both these properties and might conceivably be the 
way bars form in nature.  It is well known that spiral patterns remove angular 
momentum from stars near the inner Lindblad resonance.  The damping of the wave 
in second order perturbation theory (Kalnajs 1971; Lynden-Bell \& Kalnajs 1972; 
Mark 1974) depends on the assumption that the perturbation is too weak to trap 
the stars, but larger amplitude patterns can cause stars to be trapped.  
Simulations (Sellwood 1981; Sellwood \& Moore 1999) show that trapping in this 
way leads to bar which grows in length and strength in an episodic manner with 
each favorable pattern.  It should be noted that this process differs from that 
proposed by Lynden-Bell (1979) because it relies on rapid changes in the angular 
momenta (or actions) of stars.

Is this the way real bars are formed?  Further work is in hand to test whether 
the mechanism can account for the observed fraction of bars, the distribution of 
bar strengths and whether the resulting bars have properties consistent with 
those observed.

\section{Dynamical friction on bars}
Even though we do not understand their origin, bars allow us to impose powerful 
constraints on the dark matter (DM) content of the inner regions of galaxies.  
Weiner \etal\ (2001) show how the non-axisymmetric flow pattern driven by a bar 
can be used to determine the M/L of the bar and disk; they find, for the case of 
NGC~4123, that the stars must contribute essentially all the central attraction 
to account for the orbital speeds in the inner parts of that galaxy, and that 
the central density of DM must be low.  Their powerful argument, however, does 
not belong in this review because it involves no $N$-body simulations.

Debattista \& Sellwood (1998, 2000) use $N$-body simulations to obtain a similar 
constraint on the inner DM content of barred galaxies.  Tremaine \& Weinberg 
(1984a) argued that a bar rotating in a dense DM halo would lose angular 
momentum to the halo through dynamical friction, and Weinberg (1985), using 
perturbation theory, showed that the time scale for bar slow-down could be short 
(a few bar rotations).

There is strong, but not overwhelming, evidence to indicate that bars in real 
galaxies are fast -- \ie\ corotation lies only just beyond the end of the bar.  
Merrifield \& Kuijken (1995), Gerssen \etal\ (1999) and Debattista \& Williams 
(2001) apply the method of Tremaine \& Weinberg (1984b) to three barred 
galaxies, finding in all cases that the bar is fast.  Modeling the gas flow 
yields the pattern speed less directly; our work on NGC~4123 (above), and that 
of Lindblad \etal\ (1996) and Regan \etal\ (1997), again indicates that bars in 
real galaxies rotate rapidly.  The position of dust lanes in bars suggests that 
fast rotation is generic (van Albada \& Sanders 1982; Prendergast 1983; 
Athanassoula 1992).  Thus it appears that bars in real galaxies have not 
suffered much friction.

Our simulations confirmed Weinberg's prediction of rapid bar slow down when the 
halo has a moderate central density, unless the halo has large angular momentum 
in the same sense as the disk.  Friction decreases the bar pattern speed, 
driving the corotation point out to unacceptable distances.  The halo angular 
momentum required to avoid strong braking is unrealistically large, even when 
the halo is flattened to $b/a \simeq 0.5$ and rotation is confined to the inner 
halo only.  Bars are therefore able to maintain their observed high pattern 
speeds only if the halo has a central density low enough for the disk to provide 
most of the central attraction in the inner galaxy.

\begin{figure}[t]
\centerline{\psfig{figure=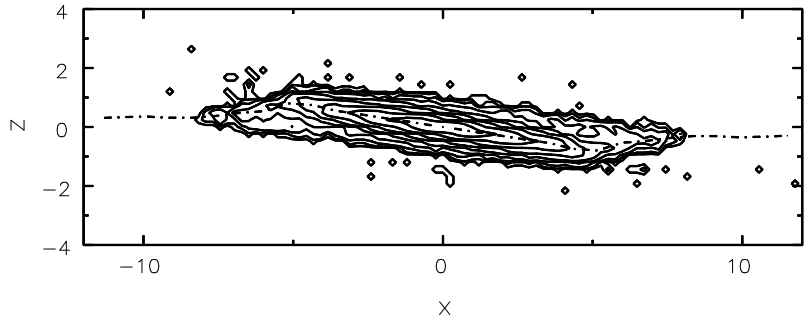,width=0.9\hsize,angle=0,clip=}}
\caption{Contours of projected disk density after about 3 orbits at $R=5$ disk 
scales.  By this time, the inner disk has tilted $\sim 10^\circ$ away from its 
original plane, which was horizontal.  The dot-dash line indicates the 
cross-section of a layer of test particles, which should trace the expected 
locus of the HI layer.}
\end{figure}

\section{Friction-induced warps}
Debattista \& Sellwood (1999) found a quite different use for dynamical friction 
between the disk and halo:  If the spin axes of the disk and of the DM halo are 
misaligned, dynamical friction will cause the angular momentum vectors to move 
closer towards alignment.  This occurs even when the halo is not significantly 
flattened by rotation.

Their simulations of this process showed that the inner disk tips more rapidly 
than the outer disk, because its higher density induces a stronger frictional 
response.  The outer disk therefore lags as the inner disk tips, creating a warp 
as shown in Figure 2.  The inner tilt is straight while the warp has a leading 
twist, just as observed (\eg\ Briggs 1990).

Note that the weak coupling between the inner and outer disk is an {\it 
advantage\/} in this mechanism, since it creates the warp, whereas it is the 
principal obstacle for would-be warp mode theories which founder because 
coupling is too weak to persuade the edge to precess at uniform rate (\eg\ 
Toomre 1983).

The warp shown in Figure 2 does not last indefinitely; it is a transient 
response when the two spin vectors find themselves out of alignment.  
Fortunately, a duration of several galaxy rotations is plenty long enough to be 
consistent with the observed prevalence of warps, as it is likely the spin 
vector of the halo changes from time to time as a result of late minor mergers 
(\eg\ Quinn \& Binney 1992), the natural consequence of any hierarchical theory 
of structure formation.  (See also the contributions by Moore and by Bullock to 
these proceedings.)

It should be noted that random motions of stars in the plane provide a 
significant additional source of stiffness for a disk, one which is ignored in 
the usual treatment of the disk as a collection of wire rings coupled only by 
gravity.  Random motion is important because adjacent mass elements of the disk 
share the same stars, and therefore cannot tip relative to each other as much as 
if gravity were the only restoring force.  Debattista \& Sellwood (1999) present 
the only calculations so far to include this effect, and show that the stiffness 
of the disk rises markedly with the level of random motion of the disk stars. 

\section*{Acknowledgments}
I would like to thank V. Debattista for comments on a draft version.  This work 
was supported by NASA LTSA grant NAG 5-6037 and by NSF grant AST-0098282.

\section*{References}
\parindent0pt
\footnotesize
\everypar{\hangindent 1cm}
\def\aap{A. \& A.}
\def\aj{Astron.\ J.}
\def\apj{Ap.\ J.}
\def\apjl{Ap.\ J.}
\def\mnras{MNRAS}

Abraham, R. G., Merrifield, M. R., Ellis, R. S., Tanvir, N. \& Brinchman, J. 
1999, \mnras, {\bf 308}, 596

Athanassoula, E. 1992, \mnras, {\bf 259}, 345

Becklin, E. E. \& Neugebauer, G. 1968, \apj, {\bf 151}, 145

Benedict, G. F., Howell, A., Jorgensen, I., Chapell, D., Kenney, J. \& Smith, B. 
J. 1998, STSCI Press Release C98

Binney, J. J., Gerhard, O. E., Stark, A. A., Bally, J. \& Uchida, K. I. 1991, 
\mnras, {\bf 252}, 210

Briggs, F. H. 1990, \apj, {\bf 352}, 15

Bunker, A., \etal\ 2000, astro-ph/0004348

Buta, R. \& Crocker, D. A. 1993, \aj, {\bf 105}, 1344

Debattista, V. P. \& Sellwood, J. A. 1998, \apjl, {\bf 493}, L5

Debattista, V. P. \& Sellwood, J. A. 1999, \apjl, {\bf 513}, L107

Debattista, V. P. \& Sellwood, J. A. 2000, \apj, {\bf 544}, 704

Debattista, V. P. \& Williams, T. B. 2001, in {\it Galaxy Disks and Disk 
Galaxies}, eds.\ J. G. Funes SJ \& E. M. Corsini, ASP Conference Series {\bf 
230}, (San Francisco: Astronomical Society of the Pacific) p~553

Friedli, D. 1994, in {\it Mass-Transfer Induced Activity in Galaxies}, ed.\ I. 
Shlosman (Cambridge: Cambridge University Press) p~268

Gerhard, O. E. \& Binney, J. 1985, \mnras, {\bf 216}, 467

Gerssen, J., Kuijken, K. \& Merrifield, M. R. 1999, \mnras, {\bf 306}, 926

Hernquist, L. \& Ostriker, J. P. 1992, \apj, {\bf 386}, 375

Hohl, F. 1971, \apj, {\bf 168}, 343

Jogee, S. \& Kenney, J. 2000, in {\it Dynamics of Galaxies From the Early 
Universe to the Present}, eds.\ F. Combes, G. A. Mamon \& V. Charmandaris, ASP 
Conference Series {\bf 197}, (San Francisco: Astronomical Society of the 
Pacific) p~193

Kalnajs, A. J. 1971, \apj, {\bf 166}, 275

Kalnajs, A. J. 1972, \apj, {\bf 175}, 63

Lindblad, P. A. B., Lindblad, P. O. \& Athanassoula, E. 1996, \aap, {\bf 313}, 
65

Lynden-Bell, D. 1979, \mnras, {\bf 187}, 101

Lynden-Bell, D. \& Kalnajs, A. J. 1972, \mnras, {\bf 157}, 1

Mark, J. W-K. 1974, \apj, {\bf 193}, 539

May, A., van Albada, T. S. \& Norman, C. A. 1985, \mnras, {\bf 214}, 131

McGlynn, T. A. 1984, \apj, {\bf 281}, 13

Merrifield, M. R. \& Kuijken, K. 1995, \mnras, {\bf 274}, 933

Merritt, D. \& Quinlan, G. 1998, \apj, {\bf 498}, 625 (MQ98)

Merritt, D. \& Ferrarase L. 2001, astro-ph/0107134

Miller, R. H., Prendergast, K. H. \& Quirk, W. J. 1970, \apj, {\bf 161}, 903

Norman, C. A., Sellwood, J. A. \& Hasan, H. 1996, \apj, {\bf 462}, 114

Prendergast, K. H. 1983, in {\it Internal Kinematics and Dynamics of Galaxies}, 
IAU Symp.\ {\bf 100}, ed E. Athanassoula (Dordrecht: Reidel) p~215

Quinn, T. \& Binney, J. 1992, \mnras, {\bf 255}, 729

Regan, M. W., Vogel, S. N. \& Teuben, P. J. 1997, \apjl, {\bf 482}, L143

Sakamoto, K., Okamura, S. K., Ishizuki, S. \& Scoville, N. Z. 1999, \apj, {\bf 
525}, 691

Schwarzschild, M. 1979, \apj, {\bf 232}, 236

Sellwood, J. A. 1981, \aap, {\bf 99}, 362

Sellwood, J. A. 1985, \mnras, {\bf 217}, 127

Sellwood, J. A. 1997, in {\it Computational Astrophysics}, eds.\ D. A. Clarke \& 
M. J. West (San Francisco: ASP Conference Series {\bf 123}), 215

Sellwood, J. A. 1999, in {\it Galaxy Dynamics -- A Rutgers Symposium}, eds.\ D. 
Merritt, J. A. Sellwood \& M. Valluri (San Francisco: ASP) {\bf 182}, p~351

Sellwood, J. A. \& Evans, N. W. 2001, \apj, {\bf 546}, 176

Sellwood, J. A. \& Moore, E. M. 1999, \apj, {\bf 510}, 125

Sellwood, J. A. \& Wilkinson, A. 1993, Rep.\ Prog.\ Phys., {\bf 56}, 173

Toomre, A. 1981, in {\it The Structure and Evolution of Normal Galaxies}, ed.\ 
S. M. Fall \& D. Lynden-Bell (Cambridge: Cambridge University Press), p.~111

Toomre, A. 1983, in {\it Internal Kinematics and Dynamics of Galaxies}, IAU 
Symposium {\bf 100}, ed E. Athanassoula (Dordrecht: Reidel) p~177

Tremaine, S. \& Weinberg, M. D. 1984a, \mnras, {\bf 209}, 729

Tremaine, S. \& Weinberg, M. D. 1984b, \apjl, {\bf 282}, L5-L7

van Albada, T. S. \& Sanders, R. H. 1982, \mnras, {\bf 201}, 303

van den Bergh, S. \etal\ 1996, \aj, {\bf 112}, 359

van den Bergh, S. \etal\ 2000, \aj, {\bf 120}, 2190

van den Bergh, S., Cohen, J. G. \& Crabbe, C. 2001, astro-ph/0104458

Weinberg, M. D. 1985, \mnras, {\bf 213}, 451

Weiner, B. J. \& Sellwood, J. A. 1999, \apj, {\bf 524}, 112

Weiner, B. J., Sellwood, J. A. \& Williams, T. B. 2001, \apj, {\bf 546}, 931


\end{document}